\documentclass[sn-mathphys-num]{sn-jnl}

\usepackage{graphicx}%
\usepackage{multirow}%
\usepackage{amsmath,amssymb,amsfonts}%
\usepackage{amsthm}%
\usepackage{mathrsfs}%
\usepackage[title]{appendix}%
\usepackage{xcolor}%
\usepackage{textcomp}%
\usepackage{manyfoot}%
\usepackage{booktabs}%
\usepackage{algorithm}%
\usepackage{algorithmicx}%
\usepackage{algpseudocode}%
\usepackage{listings}%

\usepackage{bm} 
\usepackage{lmodern}           
\usepackage[T1]{fontenc}       
\usepackage[cp1250]{inputenc}  
\usepackage{verbatim} 	       
\usepackage{setspace}

\theoremstyle{thmstyleone}%
\theoremstyle{thmstyletwo}%

\theoremstyle{thmstylethree}%

\begin{document}

\title[Article Title]{On Potentials and Complementary Potentials in One-Dimensional Nonlocal Integral Formulations}


\author*[1]{\fnm{Marko} \sur{\v{C}ana\dj{}ija}} \email{marko.canadija@riteh.uniri.hr}
\author[1]{\fnm{Ante} \sur{Skoblar}} 

\affil[1]{\orgdiv{Department of Engineering Mechanics}, \orgname{Faculty of Engineering, University of Rijeka}, \orgaddress{\street{Vukovarska 58}, \city{Rijeka}, \postcode{51000}, \country{Croatia}}}


\abstract{The present research presents potentials and complementary potentials used in the one-dimensional nonlocal integral formulations. The pure stress and the pure strain nonlocal formulations were considered. While the potential used in the strain driven formulation is well known, the complementary potential has not yet been presented in the literature. The same applies to the stress driven formulation. The equivalent formulations are obtained by resorting to the Legendre transformation, and their equivalence is proved. It is also shown that these results can be used to postulate a novel potential, i.e. a kind of mixed stress-strain potential, which is, however, as ill-conditioned as the pure strain-driven formulation. Finally, an example is given that practically confirms that the stress-driven formulations resulting from the potential and the complementary potential are equivalent.}

\footnotetext{Preprint of the published article: Čanađija, M., Skoblar, A. On potentials and complementary potentials in one-dimensional nonlocal integral formulations. Meccanica 60, 3387–3396 (2025).\url{https://doi.org/10.1007/s11012-025-01985-5}. }




\maketitle

\section{Introduction}\label{sec1}

Stress and strain driven nonlocal integral formulations are a well-researched topic nowadays. These typically rely on postulating a potential from which a specific constitutive law can be derived \cite{barretta2019variational}. In the case of the strain driven problem, existence of a strain energy function is postulated. When the strain energy is placed in the strict thermodynamic framework, more specifically in the second law of thermodynamics, the stress is then determined by differentiating this energy with respect to the strain. This theoretical framework is well established, see \cite{Coleman_Gurtin:67, Lubarda04} for a local formulation or \cite{Eringen02} for a nonlocal formulation. 

It is also well known that the roles of stress and strain can be reversed \cite{Reddy2017,Washizu1968}. In other words, it is assumed that there exists a complementary strain energy, which can then be differentiated with respect to the stress to obtain the strain. However, the exact form of such a complementary potential is not presented in the literature for the nonlocal integral formulations. The aim of this paper is to find the specific form of the complementary potential in the pure strain driven formulation. A similar procedure is also applied to the pure stress-driven problem. Furthermore, explicit integral formulations for the stress and strain calculation resulting from such potentials are presented, as well as corresponding differential problems in conjunction with the constitutive boundary conditions.

\section{Potentials and complementary potentials of nonlocal integral formulations}\label{sec2}

\subsection{Definitions of stress and strain in the presence of nonlocal effects}\label{sec2_0}

The nonlocal model considered here is based on the classical one-dimensional strain or stress field, i.e. higher-order gradients are not taken into account. The following elaborations are based on \cite{lim2015higher,Eringen02}, with slight modifications. For this purpose, consider a rod of length $L$. A potential between \textit{two} points at the longitudinal coordinates $x,\xi$ is now used in the form:
\begin{equation}\label{eq:cont_2atoms}
	F=F(\varepsilon(x),\varepsilon(\xi)),
\end{equation}
where $\varepsilon(x)$ and $\varepsilon(\xi)$ are the one-dimensional strains at the point $x$ and $\xi$, respectively. The potential at point $x$, which accounts for the long-distance effects of \textit{all} points on the rod $\xi \in \left[ 0,L \right]$, is now:
\begin{equation}\label{eq:cont_allatoms}
	\rho \psi(x) =\int_{0}^{L} F(\varepsilon(x),\varepsilon(\xi)) \mathrm{d} \xi.
\end{equation}
The total potential for the entire rod, which is the sum of the potentials at each point of the rod $x \in \left[ 0,L \right]$, is then defined by the following integral:
\begin{equation}\label{eq:cont_rod}
	U = \int_{0}^{L} \rho \psi(x) \mathrm{d} x = \int_{0}^{L} \int_{0}^{L} F(\varepsilon(x),\varepsilon(\xi)) \mathrm{d} \xi  \mathrm{d} x.
\end{equation}
For such a general setting, we note that the second law of thermodynamics for the reversible processes, i.e. in the isothermal environment and for the elastic material, and with the application of the first law of thermodynamics, assumes the global form \cite{canadija2023book}:
\begin{equation}\label{eq:2ndLaw}
	\int_{0}^{L} \left(\sigma(x) \dot{\varepsilon}(x)- \rho \dot{\psi}(x) \right) \mathrm{d} x=0,
\end{equation}
where $\sigma (x)$ is the one-dimensional stress at the point $x$. For the general form of the potential Eq.~(\ref{eq:cont_allatoms}), the time rate is:
\begin{equation}\label{eq:dot_cont_allatoms}
	\rho \dot{\psi}(x) =\int_{0}^{L} \left( \frac{\partial F}{\partial \varepsilon(x)} \dot{\varepsilon}(x) +\frac{\partial F}{\partial \varepsilon(\xi)} \dot{\varepsilon} (\xi) \right) \mathrm{d} \xi.
\end{equation}
The introduction of this time rate into Eq.~(\ref{eq:2ndLaw}) then provides:
\begin{equation}\label{eq:stress}
	\int_{0}^{L} \left(\sigma \dot{\varepsilon}- \int_{0}^{L} \left(\frac{\partial F}{\partial \varepsilon(x)} \dot{\varepsilon}(x) +\frac{\partial F}{\partial \varepsilon(\xi)} \dot{\varepsilon}(\xi) \right) \mathrm{d} \xi \right) \mathrm{d} x=0.
\end{equation}
The symmetricity of the problem can now be stated in the form:
\begin{equation}\label{eq:cont_2atomssym}
	F(\varepsilon(x),\varepsilon(\xi))=F(\varepsilon(\xi),\varepsilon(x)).
\end{equation}
This means that the coordinates $x$ and $\xi$ can be interchanged in the particular function without affecting the value of the function. This operation will be denoted by the operator $\text{sym}(\bullet)$. Note that the same applies to the first derivative of the function for the symmetric function. Use of the notation:
\begin{align}\label{eq:sym2a}
	\partial_{\varepsilon(x)} F\left(\varepsilon(x),\varepsilon(\xi)\right)=\frac{\partial F \left(\varepsilon(x),\varepsilon(\xi)\right)}{\partial \varepsilon(x)}, & \quad \partial_{\varepsilon(\xi)} F\left(\varepsilon(x),\varepsilon(\xi)\right)=  \frac{\partial F\left(\varepsilon(x),\varepsilon(\xi)\right)}{\partial \varepsilon(\xi)}
\end{align}
gives:
\begin{equation}\label{eq:sym2}
	\text{sym} \left( \partial_{\varepsilon(\xi)} F \right) =\partial_{\varepsilon(x)} F \quad  \text{sym} \left( \partial_{\varepsilon(x)} F \right) =\partial_{\varepsilon(\xi)} F,
\end{equation}
i.e. if in the first derivative with respect to $\xi$, variables $x$ and $\xi$ are interchanged, then the first derivative with respect to $x$ is obtained.

Eq.~(\ref{eq:stress}) involves double integrals, so using the new notation and taking advantage of symmetry Eq.~(\ref{eq:sym2}) it follows:
\begin{align}\label{eq:sym3}
	\int_{0}^{L} \int_{0}^{L}  	\partial_{\varepsilon(\xi)} F \dot{\varepsilon}(\xi) \mathrm{d} \xi \mathrm{d} x = \int_{0}^{L} \int_{0}^{L}  	\partial_{\varepsilon(x)} F \dot{\varepsilon}(x) \mathrm{d} \xi \mathrm{d} x = \int_{0}^{L} \int_{0}^{L}  \text{sym} \left( \partial_{\varepsilon(\xi)} F \right) \dot{\varepsilon} (x) \mathrm{d} \xi \mathrm{d} x. 
\end{align}
With this result Eq.~(\ref{eq:stress}) now can be rewritten:
\begin{equation}\label{eq:stress20}
	\int_{0}^{L} \left \lbrace \sigma(x) \dot{\varepsilon}(x) - \int_{0}^{L} \left[  \partial_{\varepsilon(x)} F \dot{\varepsilon}(x) + \text{sym} \left( \partial_{\varepsilon(\xi)} F \right) \dot{\varepsilon}(x)\right]\mathrm{d} \xi \right \rbrace \mathrm{d} x =0.
\end{equation}
With Eq.~(\ref{eq:cont_allatoms}), and by extracting $\dot{\varepsilon}(x) $ for the non-trivial solution $\dot{\varepsilon}(x) \ne 0 $, this finally provides the definition of the nonlocal stresses as:
\begin{equation}\label{eq:stress2}
	 \sigma(x) = \int_{0}^{L} \left[  \partial_{\varepsilon(x)} F  + \text{sym} \left( \partial_{\varepsilon(\xi)} F \right) \right] \mathrm{d} \xi=\rho \partial_{\varepsilon(x)} \psi + \rho \, \text{sym} \; \left( \partial_{\varepsilon(\xi)} \psi \right).
\end{equation}
Note that the similar reasoning can be used to obtain:
\begin{equation}\label{eq:strain2}
	\varepsilon(x) = \rho \partial_{\sigma(x)} \psi^{\mathrm{c}} + \rho \, \text{sym} \; \left( \partial_{\sigma(\xi)} \psi^{\mathrm{c}} \right),
\end{equation}
where $\psi^{\mathrm{c}}$ is the complementary strain energy function.

At the end, note that no specific sources of strain have been mentioned in the above explanations. For example, strain can be caused by thermal or mechanical/elastic influences:
\begin{equation}\label{eq:strainthmech}
	\varepsilon = \varepsilon_\mathrm{th}+\varepsilon_\mathrm{el}.
\end{equation}
The above statements are also valid for this case. It is also well known that the stresses are only caused by the elastic part of the strain. In order to maintain the generality of the formulation, the following explanations will therefore be based on $\varepsilon_\mathrm{el}$.

\subsection{Strain and stress driven integral formulations}\label{sec2_1}
We would now want to apply these definitions of stress and strain to a specific nonlocal constitutive law. In the nonlocal strain driven integral formulation \cite{Eringen83}, the stress is calculated as follows:
\begin{equation}\label{eq:NL_Stress}
	\sigma(x) = \int_{L} \phi ({{x}-{\xi} )}  E {\varepsilon_\mathrm{el}}({\xi}) \mathrm{d} \xi,
\end{equation}
where $E$ is a constant representing the modulus of elasticity of a homogeneous rod of length $L$. Furthermore, $L_\lambda=\lambda L$, where $\lambda$ is the nonlocal parameter and where the averaging kernel function is \cite{Eringen83}:
\begin{equation}\label{eq:kernel}
	\phi (x-\xi)=\frac{1}{2L_\lambda} \exp\left({-\frac{|x-\xi|}{L_\lambda}}\right).
\end{equation} 
Note the existence of the symmetricity $\phi (x-\xi)=\phi (\xi-x)$. 

It should be mentioned that the nonlocal parameter as introduced above is typically considered a constant value for a given material. It is emphasized that this is only an approximation. As recently shown in \cite{lal2024prediction} for the carbon and boron nitride nanotubes, the nonlocal parameter depends on the physical dimensions, the strain level and the nature of the boundary value problem. This is consistent with a previous study on carbon nanotubes \cite{Canadija17}. However, the problem is further complicated by the fact that the cross-section of a carbon nanotube is taken to be a thin ring, which is also an approximation. Such an approximation is responsible for at least part of the observed size effects, see a series of papers on molecular dynamics and machine learning \cite{canadija2021deep, kosmerl2022, canadija2024computational, canadija2024optim} and the experimental work \cite{sun2024experimental}.

The corresponding stress driven formulation is \cite{Romano2017c}:
\begin{equation}\label{eq:NL_Strain}
	\varepsilon_\mathrm{el}(x) = \int_{L} \phi ({{x}-{\xi} )}  E^{-1} {\sigma}({\xi}) \mathrm{d} \xi.
\end{equation}
Obviously, both formulations have the same underlying structure. To make the following explanations clearer, we define the following shorthand notation $s$ and $f$ for the source and output field, respectively \cite{romano2017nonlocal,barretta2019variational}. More precisely:
\begin{align}\label{eq:shorthand}
	s:=\varepsilon_\mathrm{el}, \quad f := E^{-1} \sigma \nonumber \\
	s:=\sigma, \quad f :=E \varepsilon_\mathrm{el},
\end{align}
for the strain and stress driven formulation, respectively. With this notation, both convolutions Eqs.~(\ref{eq:NL_Stress}, \ref{eq:NL_Strain}) can be written compactly as:
\begin{equation}\label{eq:convolution}
	f(x)= (\phi * s) (x) = (\phi * s)_x  = \int_{L} \phi ({{x}-{\xi} )}  s(\xi) \mathrm{d} \xi.
\end{equation}
This notation also allows that Eqs.~(\ref{eq:stress2}, \ref{eq:strain2}) to be written as follows:
\begin{equation}\label{eq:abstract}
	f(x) = \partial_{s(x)} R + \, \text{sym} \; \left( \partial_{s(\xi)} R \right), \quad
	s(x) = \partial_{f(x)} R^\mathrm{c} + \, \text{sym} \; \left( \partial_{f(\xi)} R^\mathrm{c} \right),
\end{equation}
where $R,R^\mathrm{c}$ assumes the role of the potential and the complementary potential, respectively. It is clear that this result follows from the following choice of $F$:
\begin{equation}\label{eq:Gibbs_1a0}
F(s(x),s(\xi)) = \frac{1}{2} \phi ({{x}-{\xi} )} s(x) s(\xi) ,
\end{equation}
so that the corresponding abstract potential $R$ for this specific choice of $F$ is then:
\begin{align}\label{eq:Gibbs_1a}
	R(s) = \frac{1}{2} s(x) (\phi * s)_x.
\end{align}
Evidently, Eq.~(\ref{eq:abstract})$_1$ provides the output field $f$ from Eq.~(\ref{eq:Gibbs_1a}), which is defined by Eq.~(\ref{eq:NL_Stress}) or (\ref{eq:NL_Strain}), because:
\begin{equation}\label{eq:Gibbs_1b}
	\partial_{s(x)} R  = \text{sym} (\partial_{s(\xi)} R) = \frac{1}{2} (\phi * s)_x.
\end{equation}

As shown in \cite{Romano17,canadija2023book}, the integral formulations Eq.~(\ref{eq:abstract}) above can be replaced by the differential formulation of the form:
\begin{equation}\label{eq:NL_Stress_eps}
	s=-L_\lambda^2 f^{(2)} + f.
\end{equation}
This solution requires that the following special boundary conditions, the so-called constitutive boundary conditions (CBCs), are fulfilled at the end of the domain:
\begin{equation}\label{eq:CBC2}
	f^{(1)} - \frac{1}{L_\lambda}f =  0|_{x=0}, \quad f^{(1)} + \frac{1}{L_\lambda}f =  0|_{x=L}.
\end{equation}
With this, the nonlocal integral formulation is fully defined. Now it remains to determine the complementary energy function $R^{\mathrm{c}}$.

Now we would like to find a complementary potential $R^{\mathrm{c}}$ in which the variable is substituted by the first derivative to this variable, i.e. to perform a Legendre transformation. In this way, the complementary potential is related to the abstract potential $R$ as:
\begin{equation}\label{eq:Helm_Legrende}
	R^{\mathrm{c}}=s f - R.
\end{equation}
Substituting $s$ from Eq.~(\ref{eq:NL_Stress_eps}) into Eq.~(\ref{eq:Gibbs_1a}, \ref{eq:Helm_Legrende}) gives:
\begin{equation} \label{eq:Helm_3a2}
	\begin{array}{ll}
		R^{\mathrm{c}}(f(x),f(\xi);x) &=  (-L_\lambda^2 f^{(2)}(x) +f(x))f(x) \nonumber \\ 
		&-\frac{1}{2}\int_{L} \phi ({{x}-{\xi} )} 
		(-L_\lambda^2 f^{(2)}(\xi) + f(\xi)) 
		(-L_\lambda^2 f^{(2)}(x) + f(x))  \mathrm{d} \xi, \\
	\end{array}
\end{equation}
or compactly:
\begin{equation} \label{eq:Helm_3a2a}
	\begin{array}{ll}
		R^{\mathrm{c}} =  (-L_\lambda^2 f^{(2)}  +f )f   
		-\frac{1}{2} (-L_\lambda^2 f^{(2)}  + f ) \left( \phi * (-L_\lambda^2 f^{(2)} + f)\right)_x.
	\end{array}
\end{equation}

Since the source $s$ is defined in Eq.~(\ref{eq:abstract})$_2$, it is obtained:
\begin{equation}\label{eq:Helm_3a3}
	s(x) =  -L_\lambda^2 f^{(2)}(x)+2f(x)-\int_{L} \phi ({{x}-{\xi} )} (-L_\lambda^2 f^{(2)}(\xi) + f(\xi))  \mathrm{d} \xi ,
\end{equation}
or 
\begin{equation}\label{eq:Helm_3a4}
	s(x) =  -L_\lambda^2 f^{(2)}(x)+2f(x)-\left(\phi*(-L_\lambda^2 f^{(2)} + f)\right)_x.
\end{equation}

After a straightforward transformation, the equivalent differential formulation is then:
\begin{equation}\label{eq:Stress32_strn}
	s-L_\lambda^2s^{(2)} = f+L_\lambda^4 f^{(4)}-2L_\lambda^2f^{(2)},
\end{equation}
where $s$ and $f$ are evaluated at the point $x$. This has to be accompanied by new CBCs:
\begin{align}\label{eq:CBC21}
	\left( 2f-s -L_\lambda^2 f^{(2)} \right)^{(1)} - \frac{1}{L_\lambda}\left( 2f-s -L_\lambda^2 f^{(2)} \right) =  0|_{x=0}, \nonumber \\ 
	\left( 2f-s -L_\lambda^2 f^{(2)} \right)^{(1)} + \frac{1}{L_\lambda}\left( 2f-s -L_\lambda^2 f^{(2)} \right) =  0|_{x=L}.
\end{align}
Imposing Eqs.~(\ref{eq:CBC2}) in the above CBCs as before, finally yields:
\begin{align}\label{eq:CBC2a1}
	-s^{(1)} -L_\lambda^2 f^{(3)}  - \frac{1}{L_\lambda}\left( -s -L_\lambda^2 f^{(2)} \right) =  0|_{x=0}, \nonumber \\ 
	-s^{(1)} -L_\lambda^2 f^{(3)}  + \frac{1}{L_\lambda}\left( -s -L_\lambda^2 f^{(2)} \right) =  0|_{x=L}.
\end{align}
Note that in the strain driven formulation $s=\varepsilon_\mathrm{el}$, $f=E^{-1} \sigma$, so CBCs Eq.~(\ref{eq:CBC2}) place constraints on the stresses that are too stringent for most engineering applications. For this reason, the pure strain driven formulation is considered ill-conditioned \cite{Challamel08}. The issue will be revisited in Sec. \ref{sec3}.

The equivalence of the last differential formulation to the initial one can be proved directly, i.e. without resorting to the potential. In particular, the differentiation of Eq.~(\ref{eq:NL_Stress_eps}) gives:
\begin{equation}\label{eq:equi1}
	s^{(2)}=-L_\lambda^2 f^{(4)} + f^{(2)}.
\end{equation}
This can be multiplied by $L_\lambda^2$ and used together with Eq.~(\ref{eq:NL_Stress_eps}) in Eq.~(\ref{eq:Stress32_strn}), providing:
\begin{equation}\label{eq:equi2}
	s -(-L_\lambda^4 f^{(4)} + L_\lambda^2f^{(2)}) = f+L_\lambda^4 f^{(4)}-2L_\lambda^2f^{(2)},
\end{equation}
what is identical to Eq.~(\ref{eq:NL_Stress_eps}). As for the CBCs, application of Eq.~(\ref{eq:NL_Stress_eps}) and its first derivative to CBCs Eqs.~(\ref{eq:CBC2a1}) will provide:
\begin{align}\label{eq:CBC2a11}
L_\lambda^2 f^{(3)} - f^{(1)} -L_\lambda^2 f^{(3)} - \frac{1}{L_\lambda}\left( L_\lambda^2 f^{(2)} -f -L_\lambda^2 f^{(2)} \right) =  0|_{x=0}, \nonumber \\
L_\lambda^2 f^{(3)} - f^{(1)} -L_\lambda^2 f^{(3)} + \frac{1}{L_\lambda}\left( L_\lambda^2 f^{(2)} -f -L_\lambda^2 f^{(2)} \right) =  0|_{x=L},
\end{align}
what upon simplification provides CBCs Eq.~(\ref{eq:CBC2}). Consequently, two formulations are equivalent.

For the sake of completeness, we now apply these results to the specific formulations. The application of Eq.~(\ref{eq:Stress32_strn}) yields the following equivalent differential formulations in the strain-driven case:
\begin{equation}\label{eq:Stress32_str2n}
	E(\varepsilon_\mathrm{el}-L_\lambda^2\varepsilon_\mathrm{el}^{(2)}) = \sigma+L_\lambda^4 \sigma^{(4)}-2L_\lambda^2\sigma^{(2)},
\end{equation}
accompanied by the new CBCs, Eqs.~(\ref{eq:CBC2a1}):
\begin{align}\label{eq:CBC2a12}
	-E\varepsilon_\mathrm{el}^{(1)} -L_\lambda^2 \sigma^{(3)}  - \frac{1}{L_\lambda}\left( -E\varepsilon_\mathrm{el} -L_\lambda^2 \sigma^{(2)} \right) =  0|_{x=0}, \nonumber \\ 
	-E\varepsilon_\mathrm{el}^{(1)} -L_\lambda^2 \sigma^{(3)}  + \frac{1}{L_\lambda}\left( -E\varepsilon_\mathrm{el} -L_\lambda^2 \sigma^{(2)} \right) =  0|_{x=L}.
\end{align}
Likewise, in the stress driven formulation it is:
\begin{equation}\label{eq:Stress32}
-L_\lambda^2E^{-1}\sigma^{(2)}+	E^{-1}\sigma = L_\lambda^4 \varepsilon_\mathrm{el}^{(4)}-2L_\lambda^2\varepsilon_\mathrm{el}^{(2)}+\varepsilon_\mathrm{el},
\end{equation}
and two additional CBCs:
\begin{align}\label{eq:CBC1b}
	- \sigma^{(1)} - E L_\lambda^2 \varepsilon_\mathrm{el}^{(3)}  - \frac{1}{L_\lambda}\left( - \sigma - EL_\lambda^2 \varepsilon_\mathrm{el}^{(2)} \right) =  0|_{x=0}, \nonumber \\ 
	- \sigma^{(1)} - E L_\lambda^2 \varepsilon_\mathrm{el}^{(3)}  + \frac{1}{L_\lambda}\left( - \sigma - EL_\lambda^2 \varepsilon_\mathrm{el}^{(2)} \right) =  0|_{x=L}.
\end{align}

\subsection{A remark on mixed stress - strain driven integral formulations}\label{sec2_3}
The above explanations provided both the potential and the complementary potential for the pure stress and the pure strain formulation. However, since the strain driven approach is ill-conditioned, its direct application is severely limited. For this reason, the standard solution is to introduce a local and a nonlocal part of the potential \cite{eringen1972linear,wang2016exact, barretta2018closed, barretta2019variational}. Motivated by the newly developed functions, at this point we investigate a different approach, which is a mixed stress-strain driven integral formulation. The strain can be calculated by combining Eqs.~(\ref{eq:convolution}, \ref{eq:Helm_3a4}) and using $s:=\varepsilon_\mathrm{el}, f := E^{-1} \sigma$ from Eq.~(\ref{eq:shorthand}):

\begin{align}\label{eq:new1}
	\varepsilon_\mathrm{el}(x) &= (1-\alpha) E^{-1}\left(-L_\lambda^2 \sigma^{(2)}(x)+2\sigma(x)-\int_{L} \phi ({{x}-{\xi} )} (-L_\lambda^2 \sigma^{(2)}(\xi) + \sigma(\xi))  \mathrm{d} \xi \right) \nonumber \\
	&+ \alpha\int_{L} \phi ({{x}-{\xi} )}  E^{-1} {\sigma}({\xi}) \mathrm{d} \xi
\end{align}
where $\alpha \in \left[0,1\right]$ is the mixture parameter. After a straightforward transformation it is obtained:
\begin{align}\label{eq:new2}
	\varepsilon_\mathrm{el}(x) &= (1-\alpha) E^{-1}\left(-L_\lambda^2 \sigma^{(2)}(x)+2\sigma(x)\right) \nonumber \\
	-&E^{-1}\int_{L} \phi ({{x}-{\xi} )} \left(-L_\lambda^2 (1-\alpha) \sigma^{(2)}(\xi) + (1-2\alpha)\sigma(\xi)\right)  \mathrm{d} \xi. 
\end{align}
It is clear that setting $\alpha=1$ leads to a pure stress driven formulation, while choosing $\alpha=0$ gives the ill-conditioned pure strain formulation as obtained from the complementary strain energy. The differential formulation is then:
\begin{align}\label{eq:new2a}
	E\varepsilon_\mathrm{el}  -L_\lambda^2E\varepsilon_\mathrm{el}^{(2)} =  (1-\alpha) \left(L_\lambda^4 \sigma^{(4)}-2L_\lambda^2\sigma^{(2)} \right) +\sigma. 		
\end{align}

This solution must now, as before, be accompanied by the CBCs. These follow in the same way:
\begin{align}\label{eq:novel_CBC1a}
	2(1-\alpha)\sigma^{(1)}-E\varepsilon_\mathrm{el}^{(1)} -L_\lambda^2 (1-\alpha)\sigma^{(3)} - \frac{1}{L_\lambda}\left( 2(1-\alpha)\sigma-E\varepsilon_\mathrm{el} -L_\lambda^2 (1-\alpha)\sigma^{(2)} \right) =  0|_{x=0}, \nonumber \\ 
	2(1-\alpha)\sigma^{(1)}-E\varepsilon_\mathrm{el}^{(1)} -L_\lambda^2 (1-\alpha)\sigma^{(3)} + \frac{1}{L_\lambda}\left( 2(1-\alpha)\sigma-E\varepsilon_\mathrm{el} -L_\lambda^2 (1-\alpha)\sigma^{(2)} \right) =  0|_{x=L}.
\end{align}
A further reduction can be made by the application of Eq.~(\ref{eq:CBC2}):
\begin{align}\label{eq:novel_CBC1b}
	2(1-\alpha)\sigma^{(1)} -L_\lambda^2 (1-\alpha)\sigma^{(3)} - \frac{1}{L_\lambda}\left( 2(1-\alpha)\sigma -L_\lambda^2 (1-\alpha)\sigma^{(2)} \right) =  0|_{x=0}, \nonumber \\ 
	2(1-\alpha)\sigma^{(1)} -L_\lambda^2 (1-\alpha)\sigma^{(3)} + \frac{1}{L_\lambda}\left( 2(1-\alpha)\sigma -L_\lambda^2 (1-\alpha)\sigma^{(2)} \right) =  0|_{x=L}.
\end{align}
Note that the CBSs Eq.~(\ref{eq:CBC2}) were used in deriving the part that originates from the stress driven part of the formulation, and that they still have to be fulfilled. Note also that novel CBCs place constraints on the allowable distribution of axial force. Such strict constraints actually prohibit the practical application of the method.

The second approach provides stresses and can therefore be called a mixed strain driven approach. Combining Eqs.~(\ref{eq:NL_Stress}, \ref{eq:Helm_3a4}) and the use of $s:=\sigma, f :=E \varepsilon_\mathrm{el}$ from Eq.~(\ref{eq:shorthand}) results in the following:
\begin{align}\label{eq:new3}
	\sigma(x) & = \alpha \int_{L} \phi ({{x}-{\xi} )}  E {\varepsilon_\mathrm{el}}({\xi}) \mathrm{d} \xi \nonumber \\
	&+ (1-\alpha) E \left( 2\varepsilon_\mathrm{el}(x)-L_\lambda^2 \varepsilon_\mathrm{el}^{(2)}(x)-\int_{L} \phi ({{x}-{\xi} )} (-L_\lambda^2 \varepsilon_\mathrm{el}^{(2)}(\xi) + \varepsilon_\mathrm{el}(\xi))  \mathrm{d} \xi \right),
\end{align}
or
\begin{align}\label{eq:new4}
	\sigma(x) = (1-\alpha) E \left( 2\varepsilon_\mathrm{el}(x)-L_\lambda^2 \varepsilon_\mathrm{el}^{(2)}(x)\right) -\int_{L} \phi ({{x}-{\xi} )} (-L_\lambda^2 (1-\alpha) \varepsilon_\mathrm{el}^{(2)}(\xi) + (1-2\alpha) \varepsilon_\mathrm{el}(\xi))  \mathrm{d} \xi .
\end{align}
The differential formulation is:
\begin{align}\label{eq:new4a}
	E^{-1} \sigma  -L_\lambda^2E^{-1} \sigma^{(2)} =  (1-\alpha) \left(L_\lambda^4 \varepsilon_\mathrm{el}^{(4)}-2L_\lambda^2\varepsilon_\mathrm{el}^{(2)} \right) +\varepsilon_\mathrm{el} 		
\end{align}
with CBCs:
\begin{align}\label{eq:novel_CBC2}
	2(1-\alpha)\varepsilon_\mathrm{el}^{(1)}-E^{-1}\sigma^{(1)} -L_\lambda^2 (1-\alpha)\varepsilon_\mathrm{el}^{(3)} - \frac{1}{L_\lambda}\left( 2(1-\alpha)\varepsilon_\mathrm{el}-E^{-1}\sigma -L_\lambda^2 (1-\alpha)\varepsilon_\mathrm{el}^{(2)} \right) =  0|_{x=0}, \nonumber \\ 
	2(1-\alpha)\varepsilon_\mathrm{el}^{(1)}-E^{-1}\sigma^{(1)} -L_\lambda^2 (1-\alpha)\varepsilon_\mathrm{el}^{(3)} + \frac{1}{L_\lambda}\left( 2(1-\alpha)\varepsilon_\mathrm{el}-E^{-1}\sigma -L_\lambda^2 (1-\alpha)\varepsilon_\mathrm{el}^{(2)} \right) =  0|_{x=L}.
\end{align}
Again, note that the CBCs Eq.~(\ref{eq:CBC2}) still have to be accounted for:
\begin{align}\label{eq:novel_CBC2}
	2(1-\alpha)\varepsilon_\mathrm{el}^{(1)} -L_\lambda^2 (1-\alpha)\varepsilon_\mathrm{el}^{(3)} - \frac{1}{L_\lambda}\left( 2(1-\alpha)\varepsilon_\mathrm{el}-L_\lambda^2 (1-\alpha)\varepsilon_\mathrm{el}^{(2)} \right) =  0|_{x=0}, \nonumber \\ 
	2(1-\alpha)\varepsilon_\mathrm{el}^{(1)} -L_\lambda^2 (1-\alpha)\varepsilon_\mathrm{el}^{(3)} + \frac{1}{L_\lambda}\left( 2(1-\alpha)\varepsilon_\mathrm{el}-L_\lambda^2 (1-\alpha)\varepsilon_\mathrm{el}^{(2)} \right) =  0|_{x=L}.
\end{align}
Like in the purely strain driven formulation, constraints Eq.~(\ref{eq:CBC2}) are too strict for most structural problems and therefore the second mixed stress-strain formulation is also only applicable to a very limited extent.

\section{Applications of nonlocal one-dimensional potentials and complementary potentials to rod structures}\label{sec3}
In this section, the constitutive relations derived in the Sec. \ref{sec2} are now applied to rod structures. Due to economy in space, not all constitutive models are covered. For the cases that are not covered, the same procedure can be applied to obtain a required rod formulation. Likewise, to keep the framework and notation as simple as possible, only purely elastic problems are considered, i.e. without thermal effects,

Let us consider a rod of length $L$ loaded with a distributed axial load $q(x)$, with either the displacements $u(0)$, $u(L)$ or concentrated axial forces $\mathcal{N}_0$, $\mathcal{N}_L$ given at the ends. The governing potential of such a rod includes the internal potential and the potential of the external load:
\begin{equation}\label{eq:rod1}
	\Pi^*=\int_0^{V} R^* \mathrm{d} \tilde{V} - \int_0^L q u \mathrm{d} x - \mathcal{N}_0 u(0) - \mathcal{N}_L u(L),
\end{equation}
where the role of energy $R^*$ can be taken by either the potential $R$ or the complementary potential $R_\mathrm{c}$, in which case the notation $\Pi^*=\Pi$ or $\Pi^*=\Pi^\mathrm{c}$ is used, respectively.

Further derivations also require the balance of linear momentum $N(x)=\int_A \sigma \mathrm{d} \tilde{A}$ and the relationship between the axial displacement and the axial strain $\varepsilon(x) = \partial u / \partial x$. Recall that in the rod structures the axial stress $\sigma$ and the axial strain $\varepsilon$ are constant in a cross section. The rest of the procedure is standard. Depending on the particular choice of variables, the variational problem at hand can be:
\begin{equation}\label{eq:rod2}
	u=\arg \min_u \Pi, \quad \text{or} \quad N=\arg \min_N \Pi^\mathrm{c}.
\end{equation}
Variation of the internal potential $\Pi$ now is:
\begin{equation}\label{eq:rod1a}
	\delta_u \Pi=\int_0^{L} N\delta u^{(1)} \mathrm{d}  x- \int_0^L q  \delta u \mathrm{d} x - \mathcal{N}_0 \delta u(0) - \mathcal{N}_L \delta u(L)=0,
\end{equation}
where Eq.~(\ref{eq:stress2}) and the balance of the linear momentum $N(x)=\int_A \sigma \mathrm{d} \tilde{A}=\sigma A$ were used. Carrying out the integration by parts one obtains:
\begin{equation}\label{eq:rod1b}
	\delta_u \Pi=-\int_0^{L} (N^{(1)}+q) \delta u  \mathrm{d}  x - (N(0)+\mathcal{N}_0)\delta u(0) + (N(L)-\mathcal{N}_L) \delta u(L)=0,
\end{equation}
or due to the arbitrary character of $\delta u$
\begin{equation}\label{eq:rod3}
	N^{(1)} + q =0,
\end{equation}
so that
\begin{equation}\label{eq:rod31a}
	N= - \int_{L} q \mathrm{d}x,
\end{equation}
and boundary conditions:
\begin{align}\label{eq:rod4}
	N(0)=-\mathcal{N}_0 \quad & \text{or prescribe } u(0), \nonumber \\ 
	N(L)=\mathcal{N}_L  \quad & \text{or prescribe } u(L).
\end{align}
The above differential problem is derived without recourse to a constitutive model and for this reason is generally valid in both the local and nonlocal case. Note now that the formulation described in Sec. \ref{sec2_1} remains valid here, if Eq.~(\ref{eq:shorthand}) is replaced by :
\begin{align}\label{eq:shorthand2}
	s:=\partial u / \partial x, \quad f := (EA)^{-1} N \nonumber \\
	s:=N, \quad f :=EA \partial u / \partial x,
\end{align}
for the strain-driven and  stress-driven formulation, respectively. The above notation follows from Eq.~(\ref{eq:shorthand}) by applying the application of the balance of momentum $N^{(i)}(x)=\int_A \sigma^{(i)} \mathrm{d} \tilde{A}$ and $\varepsilon=\partial u / \partial x$. The resulting formulations are listed below, whereby the boundary conditions Eq.~(\ref{eq:rod4}) must be respected in all cases.

\textbf{Strain-driven formulation and the potential $R$:}

In the strain-driven case and potential $R$, from Eq.~(\ref{eq:NL_Stress_eps}) it is:
\begin{align}\label{eq:rod7}
	N(x)&=-EA u^{(1)} +L_\lambda^2 N^{(2)},
\end{align}
providing the governing equation
\begin{equation}\label{eq:rod8}
	-EA u^{(2)} +L_\lambda^2 N^{(3)} + q =0.
\end{equation}
CBCs are obtained from Eq.~(\ref{eq:CBC2}) after integration over the cross section:
\begin{align}\label{eq:rodCBC2}
	N^{(1)}  - \frac{1}{L_\lambda}N =  0|_{x=0}, \nonumber \\ 
	N^{(1)}  + \frac{1}{L_\lambda}N =  0|_{x=L}.
\end{align}
Note that this formulation leads to paradoxical solutions when CBCs in Eqs.~(\ref{eq:rodCBC2}) are ignored. Otherwise, these CBCs are too strict for most of engineering applications and the formulation of no use in structural mechanics, see \cite{peddieson2003application, Challamel08, Romano17, Romano17b}.

\textbf{Strain-driven formulation and the complementary potential $R^\mathrm{c}$:}

Likewise, for the strain-driven case and complementary potential, from Eq.~(\ref{eq:Stress32_strn}) it is:
\begin{align}\label{eq:rod5a}
	N(x)&= EA u^{(1)}-L_\lambda^2EAu^{(3)}-L_\lambda^4 N^{(4)} +2L_\lambda^2N^{(2)}.
\end{align}
The governing equation Eq.~(\ref{eq:rod3}) is then:
\begin{align}\label{eq:rod5c}
	EA u^{(2)}-L_\lambda^2EAu^{(4)}-L_\lambda^4 N^{(5)} +2L_\lambda^2N^{(3)}+q=0 ,
\end{align}
with CBCs Eq.~(\ref{eq:CBC2a1}):
\begin{align}\label{eq:CBC2a2}
	-EAu^{(2)} -L_\lambda^2 N^{(3)}  - \frac{1}{L_\lambda}\left( -EAu^{(1)} -L_\lambda^2 N^{(2)} \right) =  0|_{x=0}, \nonumber \\ 
	-EAu^{(2)} -L_\lambda^2 N^{(3)}  + \frac{1}{L_\lambda}\left( -EAu^{(1)} -L_\lambda^2 N^{(2)} \right) =  0|_{x=L}.
\end{align}
Again, this formulation suffers from the same problems as the formulation Eqs.~(\ref{eq:rod7}-\ref{eq:rodCBC2}).

\textbf{Stress-driven formulation and the potential $R$:}

In this case from Eq.~(\ref{eq:NL_Stress_eps}) it is:
\begin{align}\label{eq:rod5b}
	N(x)& = -EA L_\lambda^2 u^{(3)} + EA u^{(1)},
\end{align}
so that Eq.~(\ref{eq:rod31a}) takes the form:
\begin{equation}\label{eq:rod6}
	-EA L_\lambda^2 u^{(3)} + EA u^{(1)} +  \int_{L} q \mathrm{d}x =0.
\end{equation}
CBCs Eq.~(\ref{eq:CBC2}) become:
\begin{equation}\label{eq:rodCBC3}
	u^{(2)} - \frac{1}{L_\lambda}u^{(1)} =  0|_{x=0}, \quad u^{(2)} + \frac{1}{L_\lambda}u^{(1)} =  0|_{x=L}.
\end{equation}

\textbf{Stress-driven formulation and the complementary potential $R^\mathrm{c}$:}

For the stress-driven formulation and the complementary potential $R^\mathrm{c}$  Eq.~(\ref{eq:Stress32_strn}) provides:
\begin{align}\label{eq:rod5}
	N(x)=L_\lambda^2 N ^{(2)}+ EA \left\lbrace L_\lambda^4 u^{(5)} - 2L_\lambda^2{u^{(3)}} + u^{(1)} \right\rbrace,
\end{align}
while CBCs Eq.~(\ref{eq:CBC2a1}) become:
\begin{align}\label{eq:rodCBC}
	- N^{(1)} - EA L_\lambda^2 u^{(4)}  - \frac{1}{L_\lambda}\left( - N - EAL_\lambda^2 u^{(3)} \right) =  0|_{x=0}, \nonumber \\ 
	- N^{(1)} - EA L_\lambda^2 u^{(4)}  + \frac{1}{L_\lambda}\left( - N - EAL_\lambda^2 u^{(3)} \right) =  0|_{x=L}.
\end{align}

\section{Example}\label{sec4}
Consider a rod of length $L$ and cross-section $A$ subjected to a continuous axial compressive load $q(x)=-q_0 \exp x$. The Young's modulus is denoted by $E$. The rod is fixed at $x=0$, i.e. $u(0)=0$, while the other end $(x=L)$ is free. Determine the distribution of the axial displacement $u(x)$ using the pure stress integral formulations.

The distribution of the axial force $N(x)$ follows as a solution of the differential equation $N^{(1)}+q=N^{(1)}-q_0 \exp x=0$, with the condition $N(0)=F$, this gives:
\begin{equation}\label{eq:ex1}
	N(x)=F+q_0(e^x-1)
\end{equation}
so that the reaction $F$ at $x=0$ follows from the boundary condition $N(L)=0$ as $F=q_0(1-e^L)$. Substituting the reaction into Eq.~(\ref{eq:ex1}) then gives the final form:
\begin{equation}\label{eq:ex2}
	N(x)=q_0(e^x-e^L).
\end{equation}
These results can now be used to calculate the axial displacements using the above constitutive models.

\begin{figure}[h]
\centering
\includegraphics[width=0.7\textwidth]{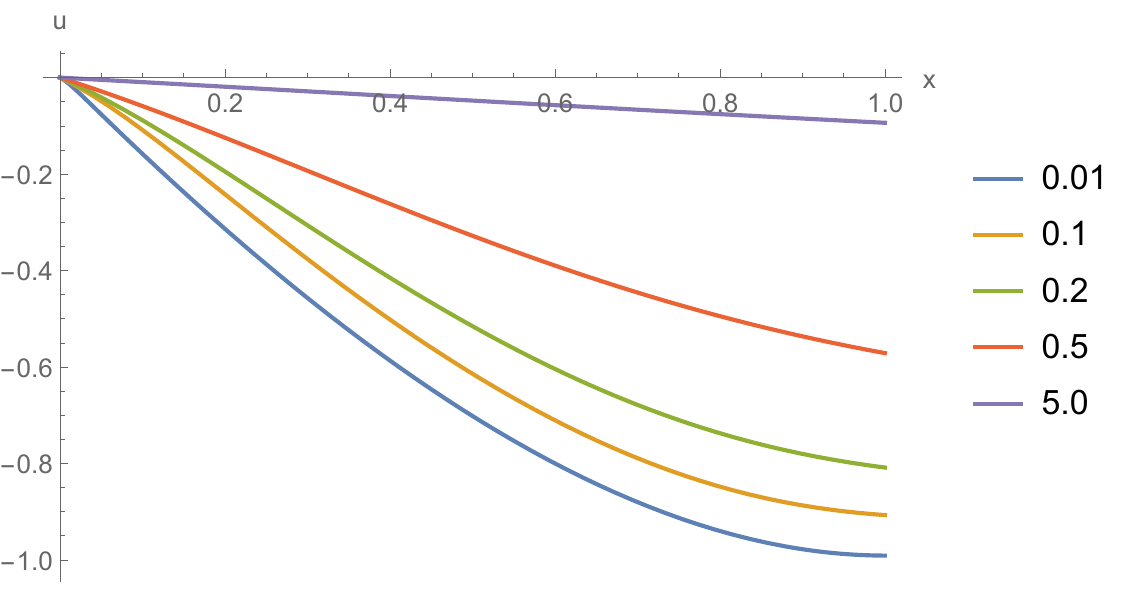}
\caption{Axial displacement along the rod for different values of the nonlocal parameter}\label{fig1}
\end{figure}

For the stress driven integral formulation that follows from the complementary potential $R^\mathrm{c}$, we solve the following problem, Eq.~(\ref{eq:rod5b}):
\begin{equation}\label{eq:ex3}
	N=AE \left(-L_\lambda^2 u^{(3)}+u^{(1)}\right)
\end{equation}
with the CBCs, Eqs.~(\ref{eq:rodCBC3}):
\begin{align}\label{eq:ex4}
	u^{(2)}  - \frac{1}{L_\lambda}u^{(1)} =  0|_{x=0}, \quad u^{(2)}  + \frac{1}{L_\lambda}u^{(1)} =  0|_{x=L},
\end{align}
and $u(0)=0$ what has to be respected in all cases. The analogous problem in the stress driven integral formulation, but based on the potential $R$ is, Eq.~(\ref{eq:rod5}):
\begin{equation}\label{eq:ex5}
	N-L_\lambda^2N^{(2)}= AE \left( L_\lambda^4 u^{(5)}-2L_\lambda^2u^{(3)}+u^{(1)} \right).
\end{equation}
Required CBCs are, Eqs.~(\ref{eq:rodCBC})::
\begin{align}\label{eq:ex6}
	-A E L_\lambda^2 u^{(4)}-N^{(1)}-\frac{1}{L_\lambda}\left(- A E L_\lambda^2 u^{(3)}-N  \right)|_{x=0} \nonumber \\
	-A E L_\lambda^2 u^{(4)}-N^{(1)}+\frac{1}{L_\lambda}\left(- A E L_\lambda^2 u^{(3)}-N  \right)|_{x=L} \nonumber \\
	u^{(2)}  - \frac{1}{L_\lambda}u^{(1)} =  0|_{x=0} \nonumber \\
	u^{(2)}  + \frac{1}{L_\lambda}u^{(1)} =  0|_{x=L} \\
\end{align}
The solution of both stress driven formulations leads to the same displacement field:
\begin{align}\label{eq:ex7}
	u(x)&=\frac {e^{-\frac{L + x}{L_\lambda}} q_0}{2 A E (L_\lambda^2-1)} 
	\left\lbrace  -2 e^{\frac {L + x + L_\lambda x}{L_\lambda}} 
	+ e^\frac{L}{L_\lambda} (L^2_\lambda- L_\lambda )   \right. \nonumber \\
	&\left. + \left(e^{L + \frac{2 x}{L_\lambda}}- e^{L\frac{x}{L_\lambda}} \right) (L_\lambda^2 + L_\lambda^3) 
	+ e^{\frac{L + x} {L_\lambda}} (2 + L_\lambda- L_\lambda^2) \right. \nonumber \\
	&\left. + e^{L + \frac{L}{L_\lambda}} (L_\lambda - L_\lambda^3)
	+ e^{\frac{L + L L_\lambda + x}{L_\lambda}} (L_\lambda^2-1) (L_\lambda - 2 x)
	\right\rbrace .
\end{align}
The distribution of axial displacements for selected values of the nonlocal parameter is shown graphically in Fig.\ref{fig1}. A clear stiffening effect can be seen when the nonlocal parameter is increased.

\section{Conclusion}\label{sec13}
The present research fills the gap in the literature regarding the specific forms of potentials and complementary potentials in nonlocal integral elasticity. In particular, the complementary potential for the pure stress and the potential for the pure strain-driven nonlocal integral formulation are well known, but the matching potentials and complementary potentials are now presented for the first time. For example, in the strain-driven formulation, the stress follows as the first derivative of the potential. In the case where the strains are required and the strain-driven formulation is used, the correct procedure for this is presented. Likewise, the equivalent differential formulation is presented together with a new pair of CBCs and the equivalence of the solutions resulting from the potential and the complementary potential is proved. 

In the end, the results presented here allow the construction of a mixed stress-strain driven integral formulation. Unfortunately, it turns out that this approach is as poorly conditioned as the pure strain integral formulation. However, although alternative formulations - two-phase local/nonlocal strain \cite{eringen1972linear,wang2016exact} or stress \cite{barretta2018closed} models, and the nonlocal strain \cite{lim2015higher,Barretta2018d} or stress \cite{barretta2019variational} gradient models were not addressed in the present research, the framework can be readily extended to include these problems as well. The same can be stated for the nonlocal thermoelastic formulations \cite{canadija2019nonlocal}. Likewise, extension to bending or torsion can be  obtained by replacing Eq.~(\ref{eq:shorthand2}) with the corresponding quantities.

\backmatter

\bmhead{Acknowledgements}
This work was supported by the University of Rijeka under project number uniri-iskusni-tehnic-23-37. All the support is gratefully acknowledged.

\section*{Declarations}
\begin{itemize}
\item Funding: his work was supported by the University of Rijeka under project number uniri-iskusni-tehnic-23-37
\item Conflict of interest/Competing interests: The authors have no competing interests to declare that are relevant to the content of this article.
\item Author contribution: Conceptualization: Marko \v{C}ana\dj{}ija; Methodology: Marko \v{C}ana\dj{}ija, Ante Skoblar; Formal analysis and investigation: Marko \v{C}ana\dj{}ija, Ante Skoblar; Writing - original draft preparation: Marko \v{C}ana\dj{}ija; Writing - review and editing: Marko \v{C}ana\dj{}ija, Ante Skoblar; Funding acquisition: Marko \v{C}ana\dj{}ija.
\end{itemize}
\noindent


\bibliographystyle{sn-mathphys-num} 

\end{document}